\documentclass[11pt]{article}

\usepackage{latexsym,amsmath,epsfig}

\DeclareFontFamily{OT1}{rsfs10}{}
\DeclareFontShape{OT1}{rsfs10}{m}{n}{ <-> rsfs10 }{}
\DeclareMathAlphabet{\mathscript}{OT1}{rsfs10}{m}{n}

\numberwithin{equation}{section}

\newcommand{\be}{\begin{equation}}
\newcommand{\ee}{\end{equation}}
\newcommand{\nn}{\nonumber}
\newcommand{\bea}{\begin{eqnarray}}
\newcommand{\eea}{\end{eqnarray}}

\newcommand{\tr}{\textrm{tr}}
\newcommand{\ns}{\normalsize}
\newcommand{\pt}{\partial}

\def\a{\alpha}
\def\b{\beta}
\def\g{\gamma}

\def\d{\delta}
\def\e{\epsilon}

\def\k{\kappa}
\def\l{\lambda}
\def\m{\mu}
\def\n{\nu}
\def\o{\omega}
\def\p{\pi}

\def\r{\rho}
\def\s{\sigma}

\def\D{\Delta}

\def\G{\Gamma}

\def\O{\Omega}

\def\cA{{\cal A}}
\def\cB{{\cal B}}

\def\cH{{\cal H}}

\def\cN{{\cal N}}

\def\cC{{\cal C}}

\def\Ib{\bar{I}}
\def\Jb{\bar{J}}
\def\Kb{\bar{K}}
\def\Lb{\bar{L}}

\def\bbar{\bar{b}}

\def\iz{i_0}
\def\jz{j_0}
\def\kz{k_0}
\def\ih{\hat{\imath}}
\def\jh{\hat{\jmath}}
\def\kh{\hat{k}}

\begin{document}


\begin{titlepage}

\vspace{-3cm}

\title{
   \hfill{\ns UPR-815T, PUPT-1809, HUB-EP-98/48, AEI-085\\}
   \hfill{\ns hep-th/9808101\\[.5cm]}
   {\LARGE Non-Standard Embedding and Five-Branes in Heterotic M--Theory}}
\author{
   Andr\'e Lukas$^{1\, 2\, 3}$, Burt A.~Ovrut$^{1\, 2}$
      \setcounter{footnote}{0}\thanks{Supported in part by a Senior 
          Alexander von Humboldt Award.}~~
   and Daniel Waldram$^4$\\[0.5cm]
   {\ns $^1$Department of Physics, University of Pennsylvania} \\
   {\ns Philadelphia, PA 19104--6396, USA}\\[0.3cm]
   {\ns $^2$Institut f\"ur Physik, Humboldt Universit\"at}\\
   {\ns Invalidenstra\ss{}e 110, 10115 Berlin, Germany}\\ [0.3cm]
   {\ns $^3$Max--Planck--Institut f\"ur Gravitationsphysik}\\
   {\ns Albert--Einstein--Institut}\\
   {\ns Schlaatzweg 1, 14473 Potsdam, Germany}\\[0.3cm] 
   {\ns $^4$Department of Physics, Joseph Henry Laboratories,}\\ 
   {\ns Princeton University, Princeton, NJ 08544, USA}}
\date{}

\maketitle

\begin{abstract}
We construct vacua of M--theory on $S^1/Z_2$ associated with Calabi--Yau
three-folds. These vacua are appropriate for compactification to $\cN=1$
supersymmetry theories in both four and five dimensions. We allow for
general $E_8\times E_8$ gauge bundles and for the presence of five-branes.
The five-branes span the four-dimensional uncompactified
space and are wrapped on holomorphic curves in the Calabi--Yau manifold.
Properties of these vacua, as well as of the resulting low-energy theories,
are discussed. We find that the low-energy gauge group is enlarged by gauge
fields that originate on the five-brane world-volumes. In addition, the 
five-branes increase the types of new $E_8\times E_8$ breaking patterns
allowed by the non-standard embedding. Characteristic features of the
low-energy theory, such as the threshold corrections to the gauge kinetic
functions, are significantly modified due to the presence of the
five-branes, as compared to the case of standard or non-standard
embeddings without five-branes. 
\end{abstract}

\thispagestyle{empty}

\end{titlepage}


\section{Introduction}

To make contact with low-energy physics, one of the central issues in
string theory has been to find vacua leading to chiral
four-dimensional theories with $\cN=1$ supersymmetry. In recent years,
the new understanding of the non-perturbative behavior of string
theory has broadened the scope for approaching these issues. First,
the M--theory paradigm of strong-weak coupling duality between the
different string theories has led to new descriptions of familiar
vacua. Second, the inclusion of brane states, that is, vacua with
non-trivial form-fields, increases the class of possible backgrounds
giving a chiral $\cN=1$ theory in four dimensions, and has raised the
possibility of gauge interactions arising from the brane world-volume
theory itself.  

In this paper, we will consider a class of eleven-dimensional
M--theory vacua based on the strongly coupled limit of the $E_8 \times
E_8$ heterotic string, as described by Ho\v rava and
Witten~\cite{hw1,hw2}. At low energy, these are compactifications of
eleven-dimensional supergravity on an $S^1/Z_2$ orbifold, with $E_8$
gauge fields at each of the two orbifold fixed planes. Following
Witten~\cite{w1}, we can further compactify on a Calabi--Yau three-fold
to give a chiral $\cN=1$ theory in four-dimensions. Most of
the discussion to date of the low-energy properties of
compactifications~\cite{hor}--\cite{BDDR} in Ho\v rava--Witten theory
has been limited to the standard embedding, where the Calabi--Yau spin
connection is embedded
in one of the $E_8$ gauge groups. Here, we will consider the general
configuration leading to $\cN=1$ supersymmetry, where, first, we allow
for general gauge bundles, and, second, include five-branes, states 
which are essentially non-perturbative in heterotic string
theory. The possibility of such generalizations was first put forward
by Witten~\cite{w1}. In ref.~\cite{stieb}, for the first time, the
gauge couplings of heterotic M--theory on non-standard embedding
Calabi--Yau manifolds without five--branes have been worked out and
compared with the results from weakly coupled heterotic string theory.
This paper also contains an interesting discussion on the effect of
five-branes on the gauge couplings. For a specific example, gauge
thresholds of non-standard embeddings in the strongly coupled limit
have also been computed in~\cite{benakli}. A toy model of gauge fields
coming from five-branes close to the orbifold planes has been presented
in~\cite{BDDR}. Finally, one notes that other limits of M--theory
leading to four-dimensional $\cN=1$ theories have recently received
attention. In particular, there has been renewed interest in the
phenomenology of type I vacua, dual to the $SO(32)$ heterotic string,
which also includes the presence of
branes~\cite{ibanez,lykken}. However, we will not consider such limits
here. 

The $\cN=1$ vacua we will discuss have the following structure. One
starts with the spacetime $M_{11}= S^1/Z_2\times X\times M_4$, where
$X$ is a Calabi--Yau three-fold and $M_4$ is flat Minkowski space.
As in the weakly coupled limit, to preserve the
four supercharges, arbitrary holomorphic $E_8$ gauge bundles over $X$
(satisfying the Donaldson--Uhlenbeck--Yau condition) are allowed on
each plane. In particular, there is no requirement that the
spin-connection of the Calabi--Yau space be embedded in the gauge connection
of one of the $E_8$ bundles. This generalization is what is meant by
non-standard embedding, and has a long history in the phenomenology of
weakly coupled strings (for early discussions see
refs.~\cite{w0,ww,gsw}). In addition, one can add five-branes, located
at points throughout the orbifold interval. The five-branes will
preserve some supersymmetry, provided the branes are wrapped
on holomorphic two-cycles within $X$ and otherwise span the flat
Minkowski space $M_4$~\cite{w1}. 

Both the gauge fields and the five-branes are magnetic sources for the
four-form field strength $G$ of the bulk supergravity, and so excite a
non-zero $G$ within the compact $S^1/Z_2\times X$ space. This has two
effects. First, since the space is compact, there can be no net
magnetic charge, for there is nowhere for the flux to ``escape''. Thus,
there is a cohomological condition that the sum of the sources must be
zero. Secondly, the non-zero form field enters the Killing spinor
equation and so, to preserve supersymmetry, the background geometry
must have a compensating distortion~\cite{w1}. This leads to a
perturbative expansion of the supersymmetric background. Such an
expansion is familiar in non-standard embeddings in the weakly
coupled heterotic string~\cite{w0,ww,gsw}. In the strongly coupled
limit, it appears even for the standard embedding. From this point of
view, the generalization to include non-standard embedding and
five-branes is very natural.

Having found the vacuum as a perturbative solution, one is then
interested in the form of the low-energy theory of the massless
excitations around this compactification. It is well known that, in
the standard embedding, to match the low-energy Newton constant and
grand unified parameters, one needs to take a Calabi--Yau manifold of size
comparable to the eleven-dimensional Planck length, with the orbifold an
order of magnitude or so larger. Thus, it is natural to consider
effective actions both in five dimensions, where only $X$ is
compactified, and four, which is appropriate to momenta below the
orbifold scale. For the standard embedding, the four-dimensional action
has been calculated to leading non-trivial order~\cite{hp,low1}. Although
the expansion is completely non-perturbative, it turns out that, to
this order, the form of the effective action is identical to the large
radius Calabi--Yau limit of the one-loop effective action calculated in the
weak limit. There are threshold corrections in the gauge couplings as
well as in the matter field K\"ahler potential. In five dimensions,
because of the non-zero mode of $G$, the theory is a form of gauged
supergravity in the bulk, coupled to gauge theories on the fixed
planes~\cite{losw1,losw2}. There is no homogeneous background solution
but, rather, the correct vacuum is a BPS domain wall solution,
supported by sources on the fixed planes and a potential in the bulk.

Calculating the modifications to the low-energy effective actions due
to non-standard embedding and five-branes will be the main point of
this paper. Our results can be summarized as follows. In section two, we
discuss the expansion of the background solution, the cohomology condition
on the five-brane and orbifold magnetic sources and the constraints on
the zeroth-order background to preserve supersymmetry. We then give the
solution to first order. Expanding in terms of eigenfunctions on the
Calabi--Yau three-fold, we show that the main contribution comes from the
massless modes. Sections three and four discuss the low-energy actions
in the case of non-standard embedding and inclusion of five-branes
respectively. This requires an analysis of the theory on the
five-brane world-volume, which is given in section 4.2. In summary, we
find 
\begin{itemize}
\item For non-standard embeddings, in the absence of five-branes, the
      five-dimensional action has the same form as in the standard
      embedding case both in the bulk 
      and on the orbifold planes. However, the values of the gauge coupling
      parameters, related to the gauging of the bulk supergravity, depend on
      the form of the non-standard embedding.
\item The non-standard embedding allows many different breaking
      patterns for the $E_8$ groups. In particular, it is no longer
      necessary that the visible sector be broken to $E_6$. Rather,
      more general gauge groups $G^{(1)}, G^{(2)}\subset E_8$ and
      corresponding gauge matter can occur on the respective orbifold
      planes.
\item In the presence of five branes, the form of the bulk
      five-dimensional action between any pair of neighboring branes
      is the same as in the case of standard embedding. The
      four-dimensional fixed-plane theories also have the same
      form and couplings to the bulk fields. However, there are additional
      four-dimensional theories, arrayed throughout the orbifold and again
      coupling to the bulk fields, which arise from the five-brane
      world-volume degrees of freedom. 
\item In the conventional picture, the five-brane worldvolume theories
      provide new hidden sectors. Generically, the theory for a single
      five-brane is $\cN=1$
      supersymmetric with $g$ $U(1)$ vector multiplets, together with a
      universal chiral multiplet and a set of chiral fields parameterizing
      the moduli space of holomorphic genus $g$ two-cycles in $X$. This gauge
      group can be enhanced when five-branes overlap or when the embedding
      of a single fivebrane degenerates. In general, the total rank of the
      gauge group remains unchanged.  
\item The presence of five-branes also allows for new types of $E_8\times
      E_8$ breaking patterns, beyond those associated with non-standard
      embeddings alone. This is because the presence of five-brane sources
      leads to a wider range of solutions satisfying the zero cohomology
      condition.  
\item Reducing to four dimensions, the effective action is modified
      with respect to the standard embedding case. For pure
      non-standard embeddings, both the gauge and K\"ahler threshold
      corrections are identical in form to those of the standard
      embedding. However, the presence of the five-branes
      significantly modifies these corrections so that, for instance,
      both $E_8$ sectors can get threshold corrections of the same sign.  
\end{itemize}
The new threshold corrections due to the five-branes have no analog in
the weakly coupled limit since, first, the branes are
non-perturbative and, second, the corrections depend on the positions
of the five-branes across the orbifold, moduli which simply do not
exist in the weakly coupled limit. Similarly, the appearance of new
gauge groups due to five-branes is a non-perturbative 
effect. Finally, we note, it appears that there is 
a constraint on the total rank of the full gauge group from orbifold
fixed planes and five-branes, which arises from positivity constraints
in the magnetic charge cohomology condition. We will discuss
this issue elsewhere~\cite{lownext}.

Let us end by summarizing our conventions. We
use coordinates $x^{I}$ with indices $I,J,K,\cdots = 0,\cdots ,9,11$ to
parameterize the full eleven-dimensional space $M_{11}$. Throughout this paper,
when we refer to orbifolds, we will work in the ``upstairs'' picture
with the orbifold $S^1/Z_2$ in the $x^{11}$-direction. We choose the range
$x^{11}\in [-\pi\rho ,\pi\rho ]$ with the endpoints being identified. The
$Z_2$ orbifold symmetry acts as $x^{11}\rightarrow -x^{11}$. Then there exist
two ten-dimensional hyperplanes fixed under the $Z_2$ symmetry which we
denote by $M_{10}^{(n)}$, $n=1,2$. Locally, they are specified by the
conditions $x^{11}=0,\pi\rho$. Barred indices
$\bar{I},\bar{J},\bar{K},\dots = 0,\dots ,9$ are used for the
ten-dimensional space orthogonal to the orbifold. We use indices
$A,B,C,\dots = 4,\dots 9$ for the Calabi--Yau space.
Holomorphic and anti-holomorphic indices on the Calabi--Yau space
are denoted by $a,b,c,\dots$ and $\bar{a},\bar{b},\bar{c},\dots$,
respectively. Indices $\m ,\n ,\r ,\dots = 0,1,2,3$ are used for the
usual four space-time coordinates. Fields will be required to have a
definite behaviour under the $Z_2$ orbifold symmetry in $D=11$. We demand
a bosonic field $\Phi$ to be even or odd; that is,
$\Phi (x^{11})=\pm\Phi (-x^{11})$. For an 11--dimensional Majorana spinor
$\Psi$ the condition is $\G_{11}\Psi (-x^{11})=\pm\Psi (x^{11})$ so that
the projection to one of the orbifold planes leads to a ten-dimensional
Majorana--Weyl spinor with definite chirality. The field content of
11--dimensional supergravity is given by a metric $g_{IJ}$, an
antisymmetric tensor field $C_{IJK}$ and the gravitino $\Psi_I$. 
While $g_{\Ib\Jb}$, $g_{11,11}$, $C_{\Ib\Jb 11}$ and $\Psi_{\Ib}$ are
$Z_2$ even, $g_{\Ib 11}$, $C_{\Ib\Jb\Kb}$ and $\Psi_{11}$ are
odd. Finally, we note that we will usually adopt the convention that
the standard model gauge fields live in the bundle on the
$M^{(1)}_{10}$ fixed plane bundle, and so refer to it as the
``observable'' sector, while the bundle on the $M^{(2)}_{10}$ plane
becomes the ``hidden'' sector.

\section{Vacua with non-standard embedding and five-branes}

In this section, we are going to construct generalized heterotic M--theory
vacua appropriate for a reduction of the theory to $\cN =1$
supergravity theories in both five and four dimensions.
To lowest order (in the sense explained below), these vacua have the usual
space-time structure $M_{11}=S^1/Z_2\times X\times M_4$ where $X$ is a
Calabi--Yau three-fold and $M_4$ is four-dimensional Minkowski space.
As compared to the vacua constructed to date, we will allow for two
generalizations. First, we will not restrict ourselves to embedding the
Calabi--Yau spin connection into a subgroup $SU(3)\subset E_8$ but,
rather, allow for general (supersymmetry preserving) gauge field
sources on the orbifold hyperplanes. Secondly, we will allow for the
presence of five-branes that stretch across $M_4$ and wrap around a
holomorphic curve in $X$. As we will see, the inclusion of five-branes
makes it much easier to satisfy the necessary constraints. Therefore,
their inclusion is essential for a complete discussion 
non-standard embeddings, and leads to a considerable increase in the
number of such vacua.

\subsection{Expansion parameters}

Before we proceed to the actual computation, let us explain
the types of corrections to the lowest order background that one expects.
For the weakly coupled heterotic string, it is well known
that non-standard embeddings lead to corrections to the Calabi--Yau
background. They can be computed perturbatively~\cite{w0,ww,gsw} as a
series in
\begin{equation}
 \e_W = \frac{\a '}{v_{10}^{1/3}}\label{ew0}
\end{equation}
where $v_{10}$ is the Calabi--Yau volume measured in terms of the
ten-dimensional Einstein frame metric. At larger string coupling,
one also gets contributions from string loops. Thus the full solution
is a double expansion involving both $\e_W$ and the string coupling constant. 

On the other hand, in the strongly coupled heterotic string, it has
been shown that, even in the case of the standard embedding, there are
corrections originating from the localization of the
gauge fields to the ten-dimensional orbifold planes~\cite{w1,low1}.
Again, these corrections can be organized in a double
expansion. However, one now uses a parameterization appropriate to the
strongly coupled theory. The 11-dimensional Ho\v rava--Witten
effective action has an expansion in $\k$, the 11-dimensional
Newton constant. For the compactification on $S^1/Z_2\times X$, there
are two other scales, the size of the orbifold interval $\p\r$ and the
volume $v$ of the Calabi--Yau threefold, each measured in the
11-dimensional metric. Solving the equations of motion and
supersymmetry conditions for the action to order $\k^{2/3}$, one finds
the correction to the background is a double expansion, linear, at
this order, in the parameter 
\begin{equation}
 \e_S = \left(\frac{\k}{4\p}\right)^{2/3}
        \frac{2\p\r}{v^{2/3}}\label{es}
\end{equation}
but to all orders in
\begin{equation}
 \e_R = \frac{v^{1/6}}{\p\r}. \label{er}
\end{equation}
It is natural to use the same expansion for the background with
non-standard embedding and the inclusion of five-branes. As we will show
explicitly, the solution to the order $\k^{2/3}$ can be obtained as an
expansion in eigenfunctions of the Calabi--Yau Laplacian. It turns out
that the zero-eigenvalue, or ``massless'', terms in this expansion are
precisely of order $\e_S$, while the massive terms are of order
$\e_R\e_S$. Therefore, although one could expect corrections to
arbitrary order in $\e_R$, to leading order in $\e_S$ only the
zeroth-order and linear terms in $\e_R$ contribute. 

Clearly, for the above expansion to be valid both
$\e_S$ and $\e_R$ should be small. Let us briefly discuss the
situation at the physical point, that is, at the values of $\k$, $v$
and $\r$ that lead to the appropriate values for the four-dimensional Newton
constant and the grand unification coupling parameter and scale. There,
both the 11--dimensional Planck length $\k^{2/9}$, as well as the
Calabi--Yau radius $v^{1/6}$,
are of the order $10^{-16}\mbox{ GeV}^{-1}$ while the orbifold radius
is an order of magnitude or so larger. Inserting this into
eq.~\eqref{es} and \eqref{er} shows that $\e_S$ is of order one~\cite{bd}
while $\e_R$ is an order of magnitude or so smaller. At the physical
point, therefore, we have 
\begin{equation}
   \e_R\ll\e_S=O(1)\; .
\end{equation}
Consequently, neglecting higher-order terms in $\e_S$ might not provide a
good approximation at the physical point. It is, however, the best one
can do at the moment given that M--theory on $S^1/Z_2$ is only known
as an effective theory to order $\k^{2/3}$. On the other hand, in
fact, higher-order terms in $\e_R$ should be strongly
suppressed and can be safely neglected. 

It is interesting to note how this strong coupling expansion is
related to the weak coupling expansion with non-standard
embedding. Writing $\e_W$ in terms of 11-dimensional quantities,
one finds 
\begin{equation}
 \e_W = \left(\frac{\k}{4\p}\right)^{2/3}\frac{1}{\p^2\r v^{1/3}}\label{ew}
\end{equation}
and hence
\begin{equation}
 \e_W=\frac{1}{2\p}\e_R^2\e_S \; .\label{ew_es}
\end{equation}
Let us try to make this relation plausible. In the weak coupling limit,
the orbifold becomes small. Hence, one expects to extract the weak coupling
part of the full background by performing an orbifold average. We recall
that the massive terms in the full background are of order $\e_R\e_S$.
In addition, we will find that those massive modes decay exponentially
as one moves away from the orbifold planes, at a rate set by the Calabi--Yau
radius $v^{1/6}$. Therefore, when performing the average, one picks up
another factor of $\e_R$ leading to $\e_R^2\e_S$ as the order of the averaged
massive terms. This is in perfect agreement with the expectation,
\eqref{ew_es}, from the weakly coupled heterotic string\footnote{There is no
such comparison for the massless modes as they correspond to trivial
integration constants on the weakly coupled side which can be absorbed
into a redefinition of the moduli. This will be explained in detail later on.}.

\subsection{Basic equations and zeroth-order background}

\begin{figure}[t]
   \centerline{\psfig{figure=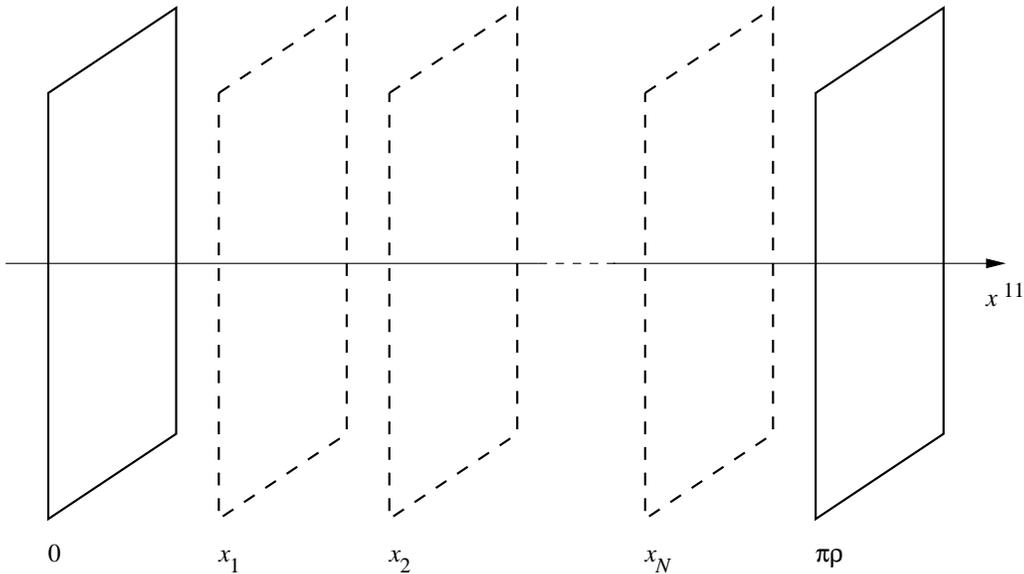,height=3in}}
   \caption{Orbifold interval with boundaries at $0$, $\p\r$ and $N$
            five-branes at $x_1,\dots ,x_N$. The mirror interval from
            $0$ to $-\p\r$ is suppressed in this diagram.}
   \label{fig1}
\end{figure}

The M--theory vacuum is given in the 11-dimensional limit by
specifying the metric $g_{IJ}$ and the three-form $C_{IJK}$ with field
strength $G_{IJKL}=24\,\partial_{[I}C_{JKL]}$. To the order
$\k^{2/3}$, the set of equations to be solved consists of the Killing spinor
equation
\begin{equation}
 \d\Psi_I = D_I\eta +\frac{\sqrt{2}}{288}
            \left(\G_{IJKLM}-8g_{IJ}\G_{KLM}\right)G^{JKLM}\eta = 0\; ,
 \label{killing}
\end{equation}
for a Majorana spinor $\eta$, the equation of motion for $G$
\begin{equation}
 D_IG^{IJKL} = 0 \label{Geom}
\end{equation}
and the Bianchi identity\footnote{Here we are using the normalization
given in ref.~\cite{hw2}. Conrad~\cite{conrad} has argued that the correct
normalization is smaller. In that case, the coefficient of the
right-hand side of the Bianchi identity~\eqref{G} and eqns.~\eqref{S}
and~\eqref{cB} below are all multiplied by $2^{-1/3}$. Furthermore, the
definition of $\e_S$ in eqn.~\eqref{es} should also be multiplied by
$2^{-1/3}$.}
\bea
 (dG)_{11\Ib\Jb\Kb\Lb} &=& 4\sqrt{2}\p\left(\frac{\k}{4\p}
                           \right)^{2/3}\left[J^{(0)}\d (x^{11})+J^{(N+1)}
                           \d (x^{11}-\p\r )+\right.\nn \\
                       &&\left.\qquad\qquad\qquad\qquad\frac{1}{2}
                         \sum_{n=1}^NJ^{(n)}(\d (x^{11}-x_n)+\d (x^{11}+x_n))
                           \right]_{\Ib\Jb\Kb\Lb}\; .\label{G}
\eea
Here the sources $J^{(0)}$ and $J^{(N+1)}$ on the orbifold planes are as
usual given by
\begin{equation}
\begin{aligned}
 J^{(0)} &= -\frac{1}{16\p^2}\left.\left(\tr F^{(1)}\wedge F^{(1)} 
      - \frac{1}{2}\tr R\wedge R\right)\right|_{x^{11}=0} \; , \\
 J^{(N+1)} &= -\frac{1}{16\p^2}\left.\left(\tr F^{(2)}\wedge F^{(2)} 
      - \frac{1}{2}\tr R\wedge R\right)\right|_{x^{11}=\p\r} \; .
\end{aligned}
\label{J} 
\end{equation}
We have also introduced $N$ additional sources $J^{(n)}$,
$n=1,\dots ,N$. They come from $N$ five-branes located at
$x^{11}=x_1,\dots ,x_N$ where $0\leq x_1\leq\dots\leq x_N\leq\p\r$
(see fig.~\ref{fig1}). Note that each five-brane at $x^{11}=x_n$ has
to be paired with a mirror five-brane at $x^{11}=-x_n$ with the same
source since the Bianchi identity must be even under the $Z_2$
orbifold symmetry. Our normalization is such that the total source of
each pair is $J^{(n)}$. The structure of these five-brane sources
will be discussed below. 
We are interested in finding solutions of these equations that preserve
$3+1$-dimensional Poincar\'e invariance and admit a Killing spinor $\eta$
corresponding to four preserved supercharges and, hence, $\cN =1$
supersymmetry in four dimensions.

The usual procedure to find such solutions is to solve the equations
perturbatively. One starts by choosing a space
$S^1/Z_2\times X\times M_4$, where $X$ is a Calabi--Yau three-fold with a
Ricci-flat metric $g_{AB}$, admitting a Killing spinor $\eta^{({\rm CY})}$.
To lowest order, the solution, denoted in the following by $(0)$, is then
given by
\begin{equation}
\begin{aligned}
 {ds^{(0)}}^2 \equiv g_{IJ}^{(0)}dx^Idx^J &= \eta_{\m\n}dx^\m dx^\n
     +g_{AB}dx^Adx^B+(dx^{11})^2 \\
 G_{IJKL}^{(0)} &= 0 \\
 \eta^{(0)} &= \eta^{({\rm CY})}\; .
\label{sol0} 
\end{aligned}
\end{equation}
Note that it is consistent, to this order, to set the antisymmetric
tensor field to zero since the sources in the Bianchi identity are
proportional to $\k^{2/3}$ and, hence, first order in $\e_S$.

One must also ensure that the theories on the orbifold planes preserve
supersymmetry. This leads to the familiar constraint, following from
the vanishing of the supersymmetry variation of the gauginos, that
\begin{equation}
\G^{\Ib\Jb}F_{\Ib\Jb}^{(1)}\eta |_{x^{11}=0}=
\G^{\Ib\Jb}F_{\Ib\Jb}^{(2)}\eta |_{x^{11}=\p\r}=0\; .\label{gau_killing}
\end{equation}
As discussed in~\cite{gsw}, this implies that each $E_8$ gauge field
is a holomorphic gauge bundle over the Calabi--Yau three-fold, satisfying the
Donaldson--Uhlenbeck--Yau condition. The holomorphicity implies that
$F^{(1)}_{AB}$ and $F^{(2)}_{AB}$ are (1,1)-forms. It follows that, since
$R_{AB}$ for a Calabi--Yau three-fold is also a (1,1)-form, the orbifold
sources $J^{(0)}$ and $J^{(N+1)}$, defined by eq.~\eqref{J}, are closed
$(2,2)$-forms.  

For the five-brane world-volume theory to be supersymmetric, the branes
must be embedded in the Calabi--Yau space in a particular way~\cite{w1}. To
preserve Lorentz invariance in $M_4$, they must span the  $3+1$-dimensional
uncompactified space. The remaining spatial dimensions must then be
wrapped on a two-cycle in the Calabi--Yau space. The condition of
supersymmetry implies that the cycle is a holomorphic
curve~\cite{w1,bbs,vb}. As we will show in section 4.2, in such a situation,
we preserve four supercharges on the five-brane worldvolume corresponding
to $\cN =1$ supersymmetry in four dimensions. Since the five-branes are
magnetic sources for $G$, they enter the right-hand side of the
Bianchi identity~\eqref{G} as source terms, which should be localized
on the five-brane world-volumes. The delta function in $x^{11}$ gives
the localization in the orbifold direction, while the four-forms
$J^{(n)}$ must give the localization of the $n$-th five-brane on the
two-cycle $\cC_2^{(n)}$. Explicitly, for any two-cycle $\cC_2$, one
can introduce a delta-function four-form $\d(\cC_2)$, defined in the
usual way, such that for any two-form $\chi$,
\begin{equation}
   \int_X \chi \wedge \d(\cC_2) = \int_{\cC_2} \chi \; ,
\end{equation}
so that $\d(\cC_2)$ is localized on $\cC_2$. In general, we would
expect that $J^{(n)}$ is proportional to $\d(\cC_2^{(n)})$. In fact,
the correct normalization of the five-brane magnetic
charge~\cite{DMW,wfq} implies that the two are equal, that is
\begin{equation}
   J^{(n)} = \d(\cC_2^{(n)}) \; .
\label{Jdef}
\end{equation}
Since the cycles are holomorphic, $J^{(n)}$, like the orbifold
sources, are closed (2,2)-forms. 

There is one further condition which the five-branes and the
fields  on the orbifold planes must satisfy. This is a cohomology condition
on the Bianchi identity~\cite{w1}. Consider integrating the
identity~\eqref{G} over a five-cycle which spans the orbifold interval
together with an arbitrary four-cycle $\cC_4$ in the Calabi--Yau three-fold.
Since $dG$ is exact, this integral must vanish. Physically this is the
statement that there can be no net charge in a compact space, since there
is nowhere for the flux to ``escape''. Performing the integral over the
orbifold, we derive, using \eqref{G}, the condition
\begin{equation}
   - \frac{1}{16\p^2}\int_{\cC_4}\tr F^{(1)}\wedge F^{(1)}
      - \frac{1}{16\p^2}\int_{\cC_4}\tr F^{(2)}\wedge F^{(2)}
      + \frac{1}{16\p^2}\int_{\cC_4}\tr R \wedge R
      + \sum_{n=1}^{N} \int_{\cC_4}J^{(n)}
      = 0\; . \label{chargecon}
\end{equation}
Hence, the net magnetic charge over $\cC_4$ is zero. Equivalently, this
implies that the sum of the sources must be cohomologically trivial, that is 
\begin{equation}
   \left[\sum_{n=0}^{N+1}J^{(n)}\right] = 0\; .\label{coh}
\end{equation}

\begin{figure}[t]
   \centerline{\psfig{figure=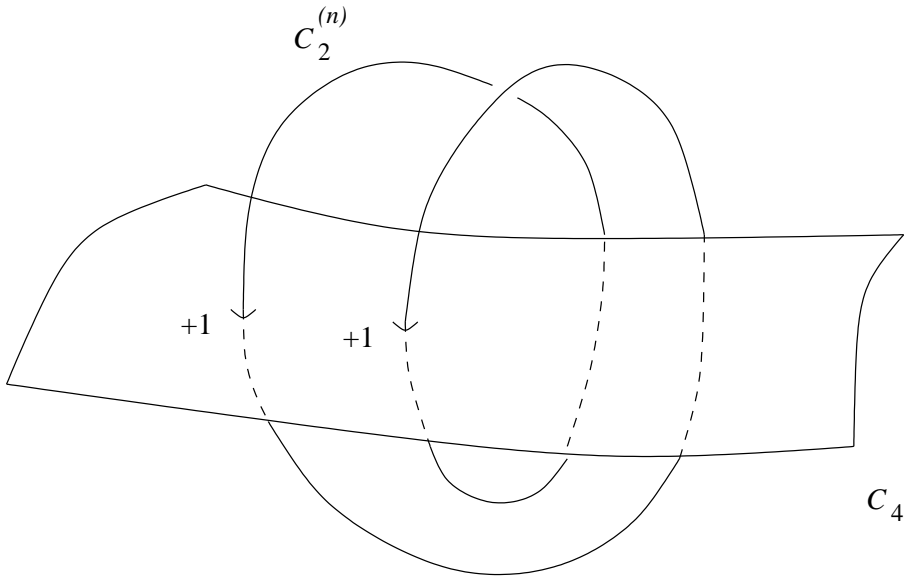,height=3in}}
   \caption{Intersection of a five-brane wrapped on the holomorphic
   cycle $\cC_2^{(n)}$ and a four-cycle $\cC_4$. In this example 
   the five-brane contributes two units of magnetic charge on $\cC_4$.}
   \label{intersect}
\end{figure}

Let us now return to the normalization of the five-brane
charges. We note that in equation~\eqref{chargecon} the first three
terms are all integers. They are topological invariants, giving the
instanton numbers (second Chern numbers) of the two $E_8$ bundles and
the instanton number (first Pontrjagin number) of the tangent bundle of
the Calabi--Yau three-fold. Hence, the above constraint shows that
$n_5(\cC_4)=\sum_{n=1}^{N}\int_{\cC_4}J^{(n)}$ must also be an
integer. In fact, with the normalization given in eqn.~\eqref{Jdef}, each
$\int_{\cC_4}J^{(n)}$ is an integer. It is also a topological
invariant, giving the intersection number~\cite{GH} of the $n$-th
brane, on the two-cycle $\cC_2^{(n)}$, with the four-cycle
$\cC_4$. This can be understood as follows (see
fig.~\ref{intersect}). 
The two cycles naturally intersect at points in
the Calabi--Yau manifold. Thus in $\cC_4$, the five-brane appears as a
set of point-like magnetic charges located at each intersection. The
net contribution of the five-brane to the magnetic charge on $\cC_4$
is then the sum of the point charges, which is precisely the
intersection number. Given the normalization of~\eqref{Jdef}, each
intersection contributes one unit of magnetic charge. We also note
that, for a holomorphic curve $\cC_4$, since $\cC_2^{(n)}$ is
holomorphic, it is a theorem~\cite{GH} that the intersection number is
always positive. This is related to the fact that only five-branes and
not anti-five-branes are allowed if we are to preserve
supersymmetry. In summary, the main point is that the normalization of
the five-brane charge is such that each five-brane intersection with
$\cC_4$ and each gauge instanton on the orbifold plane carry the same
amount of magnetic charge~\cite{DMW,wfq}. 

We can then rewrite the cohomology condition~\eqref{chargecon} on a
particular holomorphic four-cycle $\cC_4$ as 
\begin{equation}
   n_1(\cC_4) + n_2(\cC_4) + n_5(\cC_4) = n_R(\cC_4)
\label{ccon}
\end{equation}
which states that the sum of the number of instantons on the two $E_8$
bundles and the sum of the intersection numbers of each five-brane
with the four-cycle $\cC_4$, must equal the instanton number for the
Calabi--Yau tangent bundle, a number which is fixed once the
Calabi--Yau geometry is chosen.

In summary, we see that to define the zeroth-order background we must
specify the following data
\begin{itemize}
\item
a Calabi--Yau three-fold $X$,
\item
two holomorphic vector bundles over $X$, one for each fixed plane,
satisfying the Donaldson--Uhlenbeck--Yau condition. In general, there
is no constraint that these bundles correspond to the embedding of the
spin-connection in the gauge connection,
\item
a set of five-branes, each spanning the uncompactified $3+1$ dimensional
space and wrapping a holomorphic two-cycle in the Calabi--Yau space,
\item
the sum of the five-branes magnetic charges and the instanton numbers
from the gauge bundles, must equal the tangent space instanton number
of $X$, as in equation~\eqref{ccon},
\end{itemize}
We can then proceed to calculate the first-order corrections to the
background. 

\subsection{First-order background}

As an expansion in $\e_S$, we write the bulk fields and the Killing spinor as 
\begin{equation}
 \begin{aligned}
    g_{IJ} &= g_{IJ}^{(0)}+g_{IJ}^{(1)} \\
    C_{IJK} &= C_{IJK}^{(0)}+C_{IJK}^{(1)}\\
    \eta &= \eta^{(0)}+\eta^{(1)}\; .
  \end{aligned}
 \label{decomp1}
\end{equation}
where the index $(0)$ refers to the uncorrected background, given
in \eqref{sol0}, and the index $(1)$ to the corrections to first order 
in $\e_S$. 

Expanding to this order in $\e_S$, we get for the Killing spinor equation
\eqref{killing}
\begin{multline}
 \d\Psi_I = D_I^{(0)}\eta^{(1)}-\frac{1}{8}\left(D_J^{(0)}
              g_{KI}^{(1)}-D_K^{(0)}g_{JI}^{(1)}\right)
                \G^{JK}\eta^{(0)} \\
              +\frac{\sqrt{2}}{288}\left(
                \G_{IJKLM}-8g_{IJ}^{(0)}\G_{KLM}\right)
                G^{(1)JKLM}\eta^{(0)} = 0
 \label{S0}
\end{multline}
and for the equation of motion for $G$ \eqref{Geom} and the
Bianchi identity \eqref{G}
\begin{equation}
\begin{aligned}
 D_I^{(0)}G^{(1)IJKL} &= 0 \\
 (dG^{(1)})_{11\Ib\Jb\Kb\Lb} &= 4\sqrt{2}\p\left(\frac{\k}{4\p}
                                 \right)^{2/3}\left[ J^{(0)}\d (x^{11})+
                                 J^{(N+1)}\d (x^{11}-\p\r) \right.\\
                             &\qquad\qquad\qquad\qquad \left.+\frac{1}{2}
                                 \sum_{n=1}^{N}J^{(n)}(\d (x^{11}-x_n)+
                                 \d (x^{11}+x_n)) \right]_{\Ib\Jb\Kb\Lb}\;
                               . \label{S}
\end{aligned}
\end{equation}
First, we note that the only nonvanishing components of the
antisymmetric tensor $G^{(1)}$ are $G^{(1)}_{a\bar{b}c\bar{d}}$ and
$G_{a\bar{b}c11}^{(1)}$. This follows from the Bianchi identity for
$G^{(1)}$ in eq.~\eqref{S} and the fact that all sources $J^{(n)}$ are
$(2,2)$ forms. For $G^{(1)}$ of this form, the Killing spinor equation
has been analyzed in ref.~\cite{w1}. It has been shown in that paper
that the corrections first order in $\e_S$ to the metric and Killing spinor
should have the structure 
\begin{equation}
 g^{(1)}_{\m\n} = b\eta_{\m\n}\; ,\qquad g^{(1)}_{AB} = h_{AB}\; \qquad
 g^{(1)}_{11,11} = \g\; ,\qquad \eta^{(1)} = \psi\eta^{(0)}
\end{equation}
with orbifold and Calabi--Yau dependent functions $b$, $h_{AB}$, $\g$
and $\psi$. Furthermore, in \cite{w1} a consistent set of differential
equations has been derived from eq.~\eqref{S0} which determines
$b$, $h_{AB}$, $\g$ and $\psi$ in terms of $G^{(1)}$. An
explicit solution for these differential equations in terms of the dual
antisymmetric tensor $\cB$ defined by
\begin{equation}
 \cH = d\cB = *G^{(1)}
\end{equation}
was presented in ref.~\cite{low1}. In the following, we adopt the harmonic
gauge, $d^*\cB = 0$. Then, since the sources in the Bianchi
identity~\eqref{S} are $(2,2)$ forms, the only nonvanishing components
of $\cB$ are 
\begin{equation}
 \cB_{\m\n\r\s a\bar{b}} = \e_{\m\n\r\s}\cB_{a\bar{b}}
\end{equation}
with $\cB_{a\bar{b}}$ a $(1,1)$ form on the Calabi--Yau space. Using the
results of ref.~\cite{low1}, the Killing spinor equation \eqref{S0} is
solved by
\begin{equation}
\begin{aligned}
  h_{a\bbar} &= \sqrt{2}i \left( \cB_{a\bbar} 
     - \frac{1}{3}\o_{a\bbar}\cB \right) \\
  b &= \frac{\sqrt{2}}{6} \cB  \\
  \g &= -\frac{\sqrt{2}}{3} \cB \\
  \psi &= -\frac{\sqrt{2}}{24} \cB  \\
  G_{ABCD}^{(1)} &= \frac{1}{2}\e_{ABCDEF}\partial_{11}\cB^{EF}  \\
  G_{ABC11}^{(1)} &= \frac{1}{2}\e_{ABCDEF}\partial^D\cB^{EF} 
\end{aligned}
\label{sol} 
\end{equation}
where $\cB = \o^{AB}\cB_{AB}$ and $\o_{a\bar{b}}=-ig_{a\bar{b}}$ is the 
K\"ahler form. We have, therefore, explicitly expressed the complete
background in terms of the $(1,1)$ form $\cB_{a\bar{b}}$. All that
remains then is to determine this $(1,1)$ form, which can be done following
the methods given in ref.~\cite{low1}. In the harmonic gauge, which
implies
\begin{equation}
 D_A^{(0)}\cB^{AB} = 0\; ,
\end{equation}
$\cB_{AB}$ is determined from eq.~\eqref{S} by solving
\begin{multline}
  \left( \D_X + D_{11}^2 \right) \cB_{AB} = 
       4\sqrt{2}\pi\left(\frac{\k}{4\pi}\right)^{2/3}
          \left[ *_X J^{(0)} \d(x^{11}) + *_X J^{(N+1)} \d(x^{11}-\p\r)
          \right. \\ \left.
       + \frac{1}{2}\sum_{n=1}^{N} *_X J^{(n)} 
          \left( \d(x^{11}-x_n) + \d(x^{11}+x_n) \right)
          \right]_{AB}\; .\label{cB}
\end{multline}
where $\D_X$ is the Laplacian and $*_X$ the Hodge star operator on the
Calabi--Yau space. Essentially, this is the equation for a potential
between a set of charged plates positioned through the orbifold
interval at the fixed planes and the five-brane locations. The charge
is not uniform over the Calabi--Yau space. To find a solution,
following ref.~\cite{low1} we introduce eigenmodes $\p_{i\,a\bar{b}}$
of this Laplacian with eigenvalues $-\l_i^2$ so that 
\begin{equation}
  \D_X \p_{i\,a\bbar} = - \l_{i}^2 \p_{i\,a\bbar} \; .
\end{equation}
Generically, $\l_i$ is of order $v^{-1/6}$. The metric on the space of
eigenmodes
\begin{equation}
 G_{ij} = \frac{1}{2v}\int_X\p_i\wedge (*\p_j)
\end{equation}
is used to raise and lower $i$-type indices.
Particularly relevant are the massless modes with $\l_i=0$, which are
precisely the $h^{1,1}$ harmonic $(1,1)$ forms of the Calabi--Yau space.
We will also denote these harmonic $(1,1)$ forms by $\o_{iAB}$. In the
following, in order to distinguish between massless and massive modes,
we will use indices $\iz,\jz,\kz,\dots = 1,\dots,h^{1,1}$ for the former
and indices $\ih,\jh,\kh,\dots$ for the latter, while we continue
to use $i,j,k,\dots$ for all modes. Let us now expand the sources in
terms of the eigenfunctions as
\begin{equation}
 *_X J^{(n)} = \frac{1}{2v^{2/3}} \sum_i\b_{i}^{(n)}\p^i
              \label{Jexp}
\end{equation}
where
\begin{equation}
 \b_i^{(n)} = \frac{1}{v^{1/3}} \int_X\p_i\wedge J^{(n)}\; .
\end{equation}
If we introduce four-cycles $\cC_{4\iz}$ dual to the harmonic $(1,1)$
forms $\o_{\iz}$, we can write for the massless modes
\begin{equation}
 \b_{\iz}^{(n)} = \int_{\cC_{4\iz}}J^{(n)}\; .
\label{bzdef}
\end{equation}
Specifically, it follows from \eqref{J} that $\b_{\iz}^{(0)}$ and
$\b_{\iz}^{(N+1)}$ represent the instanton numbers of the gauge fields on
the orbifold planes minus half the instanton number of the tangent bundle
and, hence, are in general half-integer. Furthermore,
$\b_{\iz}^{(n)}$, $n=1,\dots ,N$ are the
five-brane charges, given by the intersection number of each
five-brane with the cycle $\cC_{4\iz}$, and are integers. Let us
also expand $\cB_{AB}$ in terms of eigenfunctions as 
\begin{equation}
 \cB_{AB} = \sum_ib_i\p_{AB}^i\label{bser}
\end{equation}
Then inserting this expansion, together with the expression~\eqref{Jexp}
for the sources, into eq.~\eqref{cB}, it is straightforward to obtain
\bea
 \left(\partial_{11}^2-\l_i^2\right)b_i &=& \frac{\sqrt{2}\e_S}{\r}
     \left[ \b_i^{(0)}\d (x^{11})
     +\b_i^{(N+1)}\d (x^{11}-\p\r )\right.\nn \\
     &&\qquad\left. + \frac{1}{2}\sum_{n=1}^{N}\b_i^{(n)}(\d (x^{11}-x_n)
       +\d (x^{11}+x_n))\right]
     \label{b}
\eea

It is then easy to solve these equation to give an explicit
solution for the massive and massless modes. We note that the size of
the sources is set by $\e_S/\r$ which, from eq.~\eqref{es},
is independent of the size of the orbifold. We first solve
eq.~\eqref{b} for the massive modes, that is, for $\l_i\neq 0$. In
terms of the normalized orbifold coordinates
\begin{equation}
 z=\frac{x^{11}}{\p\r}\; ,\qquad z_n=\frac{x_n}{\p\r}\; ,\quad n=1,\dots ,N
 \; ,
\end{equation}
$z_0=0$ and $z_{N+1}=1$, we find
\begin{multline}
 b_{\ih} = \frac{\p\e_S}{\sqrt{2}}\d_{\ih}\left[
              \left(\sum_{m=0}^n c_{\ih ,m}\b_{\ih}^m\right)
              \sinh(\d_{\ih}^{-1}|z|)
           \right. \\ \left.  
              + \left( \sum_{m=n+1}^{N+1}s_{\ih ,m}\b_{\ih}^{(m)}
                   - \frac{c_{\ih ,N+1}}{s_{\ih ,N+1}}
                     \sum_{m=0}^{N+1}c_{\ih ,m}\b_{\ih}^{(m)}
              \right)\cosh (\d_{\ih}^{-1}|z|)
           \right]
\label{massive}
\end{multline}
in the interval
\begin{equation}
 z_n\leq |z|\leq z_{n+1}\; ,\nn
\end{equation}
for fixed $n$, where $n=0,\dots ,N$. Here we have defined
\begin{equation}
 \d_{\ih} = \frac{1}{\p\r\l_{\ih}}\; ,\qquad c_{\ih ,n} = \cosh
            (\d_{\ih}^{-1}z_n)\; ,
 \qquad s_{\ih ,n} = \sinh (\d_{\ih}^{-1}z_n)\; .
\end{equation}
Note that, since the eigenvalues $\l_{\ih}$ are of order $v^{-1/6}$, the
quantities $\d_{\ih}$ defined above are of order $\e_R$. Therefore,
as already stated, the size of the massive modes is set by $\e_R\e_S$.

We now turn to the massless modes. First note that, in order to have a
solution of~\eqref{b}, we must have 
\begin{equation}
 \sum_{n=0}^{N+1}\b_{\iz}^{(n)} = 0\; .\label{coh1}
\end{equation}
However, from the definition~\eqref{bzdef}, we see that this is, of
course, none other than the cohomology condition~\eqref{coh} described
above, and so is indeed satisfied. Integrating eq.~\eqref{b} for
$\l_i=0$ we then find~\cite{low1}
\begin{equation}
 b_{\iz} = \frac{\p\e_S}{\sqrt{2}}\left[\sum_{m=0}^n
       \b_{\iz}^{(m)}(|z|-z_m)-\frac{1}{2}\sum_{m=0}^{N+1}(z_m^2-2z_m)
       \b_{\iz}^{(m)}\right]
\label{massless}
\end{equation}
in the interval
\begin{equation}
 z_n\leq |z|\leq z_{n+1}\; ,\nn
\end{equation}
for fixed $n$, where $n=0,\dots ,N$. As already discussed, the massless
modes are of order $\e_S$ and, unlike for the massive modes, no additional
factor of $\e_R$ appears.

It is important to note that there could have been an arbitrary
constant in the zero-mode solutions. However, such a constant can always be
absorbed into a redefinition of the Calabi--Yau zero modes or,
correspondingly, the low energy fields. Consequently, in the
solution~\eqref{massless} we have fixed the constant by taking the
orbifold average of the solution to be zero. This will be important later in
deriving low-energy effective actions. 

Before we discuss the implications of these equations in detail, let us
summarize our results. We have constructed heterotic M--theory backgrounds
with non-standard embeddings including the presence of
bulk five-branes. We started with a standard Calabi--Yau background
with gauge fields and five-branes to lowest order and showed that
corrections to it can be computed in a double expansion in $\e_S$ and
$\e_R$. Explicitly, we have solved the problem to linear
order in $\e_S$ and to all orders in $\e_R$. We found the massive modes
to be of order $\e_R\e_S$ while the massless modes are of order $\e_S$.
Therefore, although one could have expected corrections of arbitrary power
in $\e_R$, we only find zeroth- and first-order contributions at the 
linear level in $\e_S$. Concentrating on the leading order massless modes,
in each interval between two five-branes, $z_n\leq |z|\leq z_{n+1}$, the
massless modes vary linearly with a slope proportional to the total charge
$\sum_{m=0}^n\b_{\iz}^{(m)}$ to the left of the interval. (Note that the total
charge to the right of the interval has the same magnitude but opposite
sign due to eq.~\eqref{coh1}.) At the five-brane
locations, the linear pieces match continuously but with kinks which
lead to the delta-function sources when the second derivative is
computed. (A specific example is given in section 4.1,
see fig.~\ref{fig2}.) Similar kinks appear for the massive modes which,
however, vary in a more complicated way between each pair of
five-branes.

\section{Backgrounds without five-branes}

In this section, we will restrict the previous general solutions to the
case of pure non-standard embedding without additional five-branes and
discuss some properties of such backgrounds and the resulting low-energy
effective actions in both four and five dimensions.

\subsection{Properties of the background}

To specialize to the case without five-branes, we set $N=0$ and recall
that $z_0=0$ and $z_1=1$. Also, the vanishing cohomology condition~\eqref{coh1}
implies that we have only one independent charge
\begin{equation}
 \b_{\iz} \equiv \b_{\iz}^{(0)}=-\b_{\iz}^{(1)}\; 
\label{onebeta}
\end{equation}
per mode.
Using this information to simplify eq.~\eqref{massless}, we find for the
massless modes
\begin{equation}
 b_{\iz} = \frac{\p\e_S}{\sqrt{2}}\b_{\iz} 
    \left(|z|-\frac{1}{2}\right)\label{massless0}\; .
\end{equation}
In the same way, we obtain from eq.~\eqref{massive} for the massive modes
\begin{equation}
 b_{\ih} = \frac{\p\e_S}{\sqrt{2}}\d_{\ih}\left[ 
           (\b_{\ih}^{(0)}-\b_{\ih}^{(1)})
           \frac{\sinh (\d_{\ih}^{-1}(|z|-1/2))}{2\cosh (\d_{\ih}^{-1}/2)}-
           (\b_{\ih}^{(0)}+\b_{\ih}^{(1)})
           \frac{\cosh (\d_{\ih}^{-1} (|z|-1/2))}{2\sinh (\d_{\ih}^{-1} /2)}
           \right]\; . \label{massive0}
\end{equation}
Note that, unlike for the massless modes, here we have no relation between
the coefficients $\b_{\ih}^{(0)}$ and $\b_{\ih}^{(1)}$. Let us compare these
results to the case of the standard embedding~\cite{low1}. We see that the
massless modes solution is, in fact, completely unchanged in form from the
the standard embedding case, though the parameter $\b_{\iz}$ can be
different. This is a direct consequence of the cohomology
condition~\eqref{coh1} which, for the simple case without
five-branes, tells us that the instanton numbers on the two orbifold
planes always have to be equal and opposite. There is no similar condition
for the massive modes and we therefore expect a difference from the
standard embedding case. Indeed, the standard embedding case is obtained
from eq.~\eqref{massive0} by setting $\b_{\ih}^{(0)}+\b_{\ih}^{(1)}=0$ so that
the second term vanishes. As was noticed in ref.~\cite{low1}, the
first term in eq.~\eqref{massive0} vanishes at the middle of the interval
$z=1/2$ for all modes. Hence, for the standard embedding, at this point
the space-time background receives no correction and, in particular,
the Calabi--Yau space is undeformed. We see that the second term in
eq.~\eqref{massive0} does not share this property. Therefore, for
non-standard embeddings, there is generically no point on the orbifold
where the space-time remains uncorrected.

Furthermore, we see that the massive modes depend on the combination
$\d_{\ih}^{-1}z$ only. Therefore, in terms of the normalized orbifold
coordinate $z$ (the orbifold coordinate $x^{11}$), the massive modes
indeed fall off exponentially with a scale set by $\d_{\ih}$ (by
$v^{1/6}$). In fact, as might be expected, we see that this part of
the solution is essentially independent of the size of the
orbifold. Averaging the above expression for the massive modes over
the orbifold, one should pick up the corresponding weak coupling
correction. Clearly, as a consequence of the exponential fall-off, 
the averaging procedure leads to an additional suppression by $\e_R$.
Given that the order of a heavy mode is $\e_R\e_S$, we conclude that
its average is of the order $\e_R^2\e_S$. According to eq.~\eqref{ew_es},
this is just $\e_W$ and, hence, the expected weak coupling expansion parameter.

\subsection{Low-energy effective actions}

What are the implications of the above results for the low-energy effective
action? Since the orbifold is expected to be larger than the Calabi--Yau
radius, it is natural to first reduce to a five-dimensional effective
theory consisting of the usual $3+1$ space-time dimensions and
the orbifold and, subsequently, reduce this theory further down to four
dimensions. First, we should explain how a background
appropriate for a reduction to $\cN =1$ supersymmetry in four dimensions
can be used to derive a sensible $\cN =1$ theory in five
dimensions~\cite{low1}\footnote{By $\cN=1$ in five dimensions we mean
a theory with eight supercharges. In four dimensions, $\cN=1$ means a
theory with four supercharges.}. The point is that, as we have seen, the
background can be split into massless and massive eigenmodes. Reducing
from eleven to five dimensions on an undeformed Calabi--Yau
background, these correspond to massless moduli fields and heavy
Kaluza--Klein modes. Working to linear order in $\e_S$, the heavy
modes completely decouple from the massless modes and so can
essentially be dropped. The background then appears as a particular
solution to the five-dimensional effective action, where the moduli
depend non-trivially on the orbifold direction. Thus, in summary, to
derive the correct five-dimensional action, we need only keep the
massless modes in a reduction on an undeformed Calabi--Yau space. However,
a similar procedure is not possible for the topologically non-trivial
components $G^{(1)}_{ABCD}$ of the antisymmetric tensor field strength. Such a
configuration of the internal field strength is not a modulus, but
rather a non-zero mode. As a consequence, the proper five-dimensional
theory is obtained as a reduction on an undeformed Calabi--Yau background
but including non-zero modes for $G$. It is these non-zero modes which
introduce all the interesting structure into the theory, notably, that
in the bulk we have a gauged supergravity and that the theory admits no
homogeneous vacuum. In the case at hand, the precise structure of the
non-zero mode can be directly read off from the background as
presented.

Let us now briefly review the results of such a reduction for the
standard embedding as presented
in ref.~\cite{losw1,losw2}. It was found that the five-dimensional
effective action consists of a gauged $\cN =1$ bulk supergravity
theory with $h^{1,1}-1$ vector multiplets and $h^{2,1}+1$
hypermultiplets coupled to four-dimensional $\cN =1$ boundary
theories. The field content of the orbifold plane at $x^{11}=0$
consists of an $E_6$ gauge multiplet and $h^{1,1}$ and $h^{2,1}$ 
chiral multiplets, while the plane at $x^{11}=\p\r$ carries $E_8$
gauge multiplets only. The gauging of the bulk supergravity is with
respect to a $U(1)$ isometry in the universal hypermultiplet coset 
space with the gauge field being a certain linear combination of the
graviphoton and the vector fields in the vector multiplets. The gauging
also leads to a bulk potential for the $(1,1)$ moduli.
In addition, there are potentials for the $(1,1)$ moduli confined to the
orbifold planes which have opposite strength. As we have mentioned,
the characteristic features of this theory, such as the gauging and
the existence of the potentials, can be traced back to the existence
of the non-zero mode. Furthermore, the vacuum solution of this
five-dimensional theory, appropriate for a reduction to four
dimensions, was found to be a double BPS domain wall with the two
worldvolumes stretched across the orbifold planes.  

Which of the above features generalize to non-standard embeddings?
The spectrum of zero mode fields in the bulk will, of course, be unchanged.
Due to the nonstandard embedding, we can have more general gauge multiplets
with groups $G^{(1)}, G^{(2)}\subset E_8$ on the orbifold planes and
also corresponding observable and hidden sector matter transforming
under these groups. We are interested in the effective action up to
linear order in $\e_S$. It is clear that, as above, to this order, the
massive part of the background completely decouples from the low-energy
effective action since the massless and massive eigenfunction on the
Calabi--Yau space are orthogonal~\cite{low1}. Hence, the 
form of the effective action to linear order in $\e_S$ is completely
determined by the massless part of the background. On the other hand,
due to the cohomology condition~\eqref{coh1}, the form of the massless
part of the background corrections is same as in the standard
embedding case, as we have just shown. Hence, in deriving the
five-dimensional effective action for non-standard
embedding, we use the same non-zero mode in the reduction as for the
standard embedding. This will lead to gauging and bulk and boundary
potentials exactly as in the standard embedding case. 

Let us explain these last facts in some more detail. First, we
identify the non-zero mode of $G$ in the case of non-standard
embedding. Inserting the mode~\eqref{massless0} into the expansion for
$\cB_{AB}$, eq.~\eqref{bser}, we can use eq.~\eqref{sol} to compute
the four-form field strength $G^{(1)}$. While the massless part of
$G^{(1)}_{ABC11}$ vanishes, we find for the massless part of
$G^{(1)}_{ABCD}$
\begin{equation}
 G^{(1)} = \frac{1}{2V}*\o_{\iz}\a^{\iz}\e (x^{11})\label{nonzero}
\end{equation}
where $V$ is the Calabi--Yau volume modulus defined by
\begin{equation}
 V=\frac{1}{2\p\r v}\int_{X\times S^1/Z_2}\sqrt{^6g}\label{V}
\end{equation}
and we have introduced the parameter
\begin{equation}
 \a_{i_0} = \frac{\sqrt{2}\e_S}{\r}\b_{i_0}\; .\label{alpha}
\end{equation}
to conform with the notation of \cite{losw1,losw2}. Furthermore, $\e (x^{11})$
is the stepfunction which is $+1$ for positive $x^{11}$ and $-1$ otherwise. 
Eq.~\eqref{nonzero} is precisely the non-zero mode we have mentioned
above. Note that $V$ measures the orbifold average of the Calabi--Yau
volume in units of $v$. In general, the parameters $\a_{i_0}$ depend
on the choice of both the tangent and the gauge bundles. Explicitly,
from eqs.~\eqref{J}, \eqref{bzdef} and the cohomology
condition~\eqref{coh1}, we have, for general embeddings, 
\begin{equation}
\begin{aligned}
    \a_{i_0} &= - \frac{\e_S}{8\sqrt{2}\p^2\r}\int_{\cC_{4i_0}}\left(
          \tr F^{(1)}\wedge F^{(1)} - \frac{1}{2}\tr R\wedge R\right) \\
             &= \frac{\e_S}{8\sqrt{2}\p^2\r}\int_{\cC_{4i_0}}\left(
          \tr F^{(2)}\wedge F^{(2)} - \frac{1}{2}\tr R\wedge R\right)\; .
\end{aligned}
\end{equation}
In the case of the standard embedding, the tangent bundle and one of the
$E_8$ gauge bundles are identified, while the other gauge bundle is
taken to be trivial, so that this reduces to 
\begin{equation}
 \a_{i_0} = - \frac{\e_S}{16\sqrt{2}\p^2\r}\int_{\cC_{4i_0}}\tr R\wedge R\; .
\end{equation}
This is the relation given in ref.~\cite{losw2}. The point is that the
expression for the non-zero mode~\eqref{nonzero} has the same form for
both standard and non-standard embeddings. All that changes are the
values of the parameters $\a_{i_0}$.  

Now let us demonstrate how the gauging of the bulk supergravity arises
in the case of non-standard embedding. Consider the five-dimensional
three-form zero--mode $C_5$, with field strength $G_5$, and the part of the
11--dimensional three-form that leads to the $h^{1,1}$ vector fields
$\cA^{\iz}$, namely $C=\cA^{\iz}\wedge\o_{\iz}$. Inserting these two
fields, together with the non-zero mode~\eqref{nonzero}, into the
Chern--Simons term in the eleven-dimensional supergravity
action~\cite{losw2} leads to 
\begin{equation}
 \int_{M_{11}}C\wedge G\wedge G \sim 
     \int_{M_5}\e (x^{11})\a_{\iz}\cA^{\iz}\wedge G_5\; .
 \label{gauging}
\end{equation}
The three-form $C_5$ can be dualized to a scalar in five dimensions,
which becomes one of the four scalars $q^u$ in the universal
hypermultiplet. Then, the above term directly causes the gauging of
the isometry in the hypermultiplet coset space that corresponds to the
axionic shift in the dual scalar. The gauging is with respect to the
linear combination $\a_{\iz}\cA^{\iz}$. Explicitly, we
find~\cite{losw2} that the universal hypermultiplet kinetic term is of
the form
\begin{equation}
   \int_{M_5} \sqrt{-g} h_{uv} D_\a q^u D^\a q^v
\end{equation}
with the covariant derivative 
\begin{equation}
   D_\a q^u = \pt_\a q^u + \e (x^{11}) \a_{i_0}\cA^{i_0}_\a k^u
\end{equation}
where $k^u$ is a Killing vector in the hypermultiplet sigma-model
manifold, pointing in the direction of the axionic shift. We see that,
since the non-zero mode~\eqref{nonzero} had the same form for both
standard and non-standard embeddings, the gauging of the supergravity
also has the same form. The only difference is in the values of the
charges $\a_{i_0}$. 

Similarly, the bulk potential should have the same form in the
standard and non-standard embedding cases. Inserting the
non-zero mode~\eqref{nonzero} into the kinetic term $G\wedge *G$ of
the four-form field strength in the eleven-dimensional supergravity
action leads to a bulk potential for the volume modulus $V$ and the
other $(1,1)$ moduli. More precisely, 
one finds
\begin{equation}
 \int_{M_{11}}G^{(1)}\wedge *G^{(1)} \sim \int_{M_5}\sqrt{-g}
   V^{-2}\a_{i_0}\a_{j_0}\tilde{G}^{i_0j_0}
\label{potential}
\end{equation}
where
\begin{equation}
 \tilde{G}_{i_0j_0}=V^{2/3}G_{i_0j_0}
\end{equation}
is a renormalized metric that depends on the Calabi--Yau shape moduli
(see ref.~\cite{losw2} for details). Note that it follows from
supersymmetry that such a potential must arise when an isometry of the
universal hypermultiplet sigma-model manifold is gauged. 

The potentials on the orbifold planes arise from the ten-dimensional
actions on the planes, with the internal gauge fields and
curvature inserted. Using identities of the form 
\begin{equation}
   \int_X \o\wedge \tr R \wedge R \sim \int_X \sqrt{-g}\tr R^2
\end{equation}
we find
\begin{equation}
 \sum_{n=1}^2\int_{M_{10}^{(n)}}\sqrt{-g}\left(\tr (F^{(n)})^2-\frac{1}{2}
     \tr R^2\right)
  \sim \int_{M^{(1)}_4}\sqrt{-g}V^{-1}\a_{i_0} b^{i_0} 
       -  \int_{M^{(2)}_4}\sqrt{-g}V^{-1}\a_{i_0} b^{i_0}
\label{YM}
\end{equation}
where $b^{i_0}$ are the K\"ahler shape moduli defined by the expansion
of the K\"ahler form $\o=V^{1/3}b^{i_0}\o_{i_0}$. 
As for the standard embedding case, the potentials come out with
opposite strength, again a consequence of the cohomology
condition~\eqref{onebeta}, $\b_{i_0}^{(0)}=-\b_{i_0}^{(1)}$.

In summary, we conclude that the five-dimensional effective
action derived in ref.~\cite{losw1,losw2} for the standard embedding
is, in fact, much more general and applies, with appropriate adjustment
of the boundary field content and the charges $\a_{i_0}$, to any
Calabi--Yau-based non-standard embedding without additional
five-branes. Furthermore, the double domain wall vacuum solution of
the five-dimensional theory is unchanged, since it does not depend on
the field content on the orbifold planes. 

The four-dimensional theory is
obtained as a reduction on this domain wall. Hence, the
four-dimensional effective action will be unchanged in the case of
non-standard embeddings without five-branes, except for the possibility of more
general gauge groups and matter multiplets. One further new feature, 
in the case of non-standard embedding, is the possibility 
of gauge matter on the hidden orbifold plane. In this case,
the threshold-like correction to the matter part of the K\"ahler
potential will be different for observable and hidden sectors in the
same way the gauge kinetic functions of the two sectors differ. 

To be more concrete, let us consider the universal case with moduli
$S$ and $T$, gauge fields of $G^{(1)}\times G^{(2)}\subset E_8\times
E_8$ and corresponding gauge matter $C^{(1)}$ and $C^{(2)}$,
transforming under $G^{(1)}$ and $G^{(2)}$, respectively. Then, we have
for the K\"ahler potential and the gauge kinetic functions
\bea
 K &=& -\log (S+\bar{S})-3\log(T+\bar{T})+Z_1|C^{(1)}|^2+Z_2|C^{(2)}|^2\nn \\
 Z_1 &=& \frac{3}{T+\bar{T}}+\frac{\p\e_S\b}{S+\bar{S}}\nn \\
 Z_2 &=& \frac{3}{T+\bar{T}}-\frac{\p\e_S\b}{S+\bar{S}} 
    \label{gkf}\\
 f^{(1)} &=&S+\p\e_S\b T\nn \\
 f^{(2)} &=&S-\p\e_S\b T\nn\; .
\eea
where $\b$ is the single instanton charge, of the type defined in
eqn.~\eqref{onebeta}, corresponding to the universal K\"ahler deformation.
For vacua based on the standard embedding, it was pointed out in
ref.~\cite{w1} that, if $\b>0$ so that the smaller of the two
couplings corresponds to the observable sector, then, fitting this to
the grand unification coupling, the larger coupling is of order one at the
``physical'' point. Hence, gaugino condensation in the hidden sector
appears to be a likely scenario. We have just shown that, in fact, this
statement continues to apply to all Calabi--Yau based non-standard
embedding vacua without additional bulk five-branes, provided $\b>0$,
since the gauge kinetic functions are completely unchanged. Gaugino
condensation, therefore, appears to be a generic possibility for such
vacua. 

\section{Backgrounds with five-branes}

Let us now turn to the much more interesting case of non-standard
embeddings with five-branes in the bulk. We will concentrate on the
massless modes, since, as above, it is these modes which will determine the
low-energy action.  

\subsection{Properties of the background}

\begin{figure}[t]
   \centerline{\psfig{figure=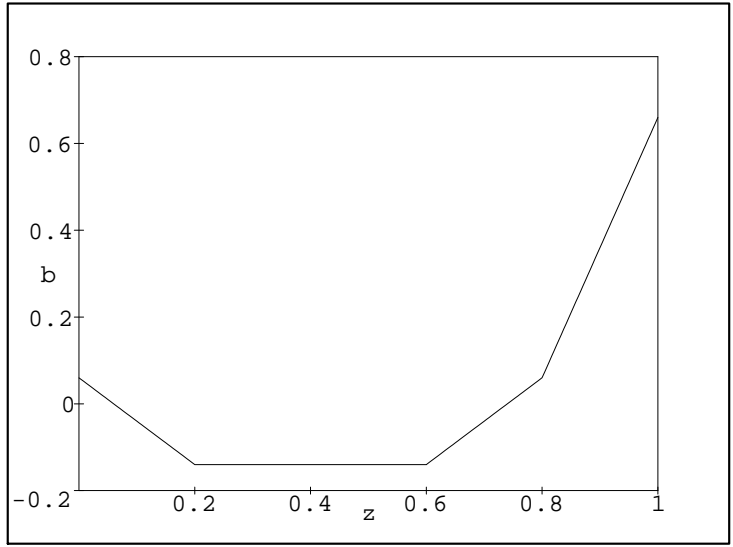,height=3in}}
   \caption{Orbifold dependence of a massless mode
   $(\sqrt{2}/\p\e_S)b$ for four five-branes at
   $(z_1,z_2,z_3,z_4)=(0.2,0.6,0.8,0.8)$ with charges
   $(\b^{(1)},\b^{(2)},\b^{(3)},\b^{(4)})=(1,1,1,1)$ and 
   instanton numbers $(\b^{(0)},\b^{(4)})=(-1,-3)$.} 
   \label{fig2}
\end{figure}

The general solution~\eqref{massless} for the massless modes shows a linear
behaviour for each interval between two five-branes. The slope, however,
varies from interval to interval in a way controlled by the five-brane
charges. The same statement applies to the variation of geometrical
quantities, like the Calabi--Yau volume, across the orbifold. Let us consider
an example for a certain massless mode $b$. Four five-branes with charges
$(\b^{(1)},\b^{(2)},\b^{(3)},\b^{(4)})=(1,1,1,1)$ are positioned at
$(z_1,z_2,z_3,z_4)=(0.2,0.6,0.8,0.8)$. Note that the third and fourth
five-brane are coincident. The instanton numbers on the orbifold
planes are chosen to be $(\b^{(0)},\b^{(4)})=(-1,-3)$. Note that the
total charge sums up to zero as required by the cohomology
constraint~\eqref{coh1}. The orbifold dependence of
$(\sqrt{2}/\p\e_S)b$ is depicted in fig.~\ref{fig2}. 
It is clear that the additional five-brane charges introduce much more
freedom as compared to the case without five-branes. For example, while
in the latter case one always has $b(0)=-b(1)$ leading to equal, but
opposite, gauge threshold corrections, the example in fig.~\ref{fig2}
shows that $b(0),b(1)>0$ is possible. One, therefore, expects the thresholds
in the low-energy gauge kinetic functions to change. This will be
analyzed in a moment. Another interesting phenomenon in the above example
is that the mode is constant between the first and second five-brane.
This is a direct consequence of our choice of the charges which sum up
to zero both to the left and the right of this interval. If such a
property is arranged for all massless modes, the Calabi--Yau volume
remains exactly constant throughout this interval.

\subsection{Five-branes on Calabi--Yau two-cycles}

The inclusion of five-branes not only generalizes the types of
background one can consider, but also introduces new degrees of
freedom into the theory, namely, the dynamical fields on the
five-branes themselves. In this section, we will consider what
low-energy fields survive on one of the five-branes when it is wrapped
around a two-cycle in the Calabi--Yau three-fold. 

In general, the fields on a single five-brane are as
follows~\cite{GT,KM}. The 
simplest are the bosonic coordinates $X^I$ describing the embedding of
the brane into 11-dimensional spacetime. The additional bosonic
field is a world-volume two-form potential $B$ with field strength $H=dB$
satisfying a generalized self-duality condition. For small
fluctuations, the duality condition simplifies to the conventional
constraint $H=*H$. These degrees of freedom are paired with spacetime
fermions $\theta$, leading to a Green--Schwarz type action, with
manifest spacetime supersymmetry and local
kappa-symmetry~\cite{BLNPST,APPS}. (As usual, including the self-dual
field in the action is difficult, but is possible by either including
an auxiliary field or abandoning a covariant formulation.) For a
five-brane in flat space, one can choose a gauge such that the
dynamical fields fall into a six-dimensional massless tensor multiplet
with $(0,2)$ supersymmetry on the brane
world-volume~\cite{Kallosh,CKvP}. The multiplet has five scalars
describing the motion in directions transverse to the five-brane,
together with the self-dual tensor $H$. 

For a five-brane embedded in $S^1/Z_2\times X \times M_4$, to preserve
Lorentz invariance in $M_4$, $3+1$ dimensions of the five-brane must
be left uncompactified. The remaining two spatial dimensions are then
wrapped on a two-cycle of the Calabi--Yau three-fold. To preserve
supersymmetry, the two-cycle must be a holomorphic
curve~\cite{w1,bbs,vb}. Thus, from the point of view of a
five-dimensional effective theory on $S^1/Z_2\times M_4$, since two of
the five-brane directions are compactified, it appears as a flat
three-brane (or equivalently domain wall) located at some point
$x^{11}=x$ on the orbifold. Thus, at low energy, the degrees of
freedom on the brane must fall into four-dimensional supersymmetric
multiplets.

\vspace{0.4cm}

An important question is how much supersymmetry is preserved in the
low-energy theory. One way to address this problem is directly from
the symmetries of the Green--Schwarz action, following the discussion
for similar brane configurations in~\cite{bbs}. Locally, the
11-dimensional spacetime $S^1/Z_2\times X\times M_4$ admits
eight independent Killing spinors $\eta$, so should be described by a
theory with eight supercharges. (Globally, only half of the spinors
survive the non-local orbifold quotienting condition
$\G_{11}\eta(-x^{11})=\eta(x^{11})$, so that, for instance, the
eleven-dimensional bulk fields lead to $\cN=1$, not $\cN=2$,
supergravity in four dimensions.) The Green--Schwarz form of the
five-brane action is then invariant under supertranslations generated
by $\eta$, as well as local kappa-transformations. In general the
fermion fields $\theta$ transform as (see for instance
ref.~\cite{CKvP})
\begin{equation}
   \d\theta = \eta + P_+\k
\end{equation}
where $P_+$ is a projection operator. If the brane configuration is
purely bosonic then $\theta=0$ and the variation of the bosonic fields
is identically zero. Furthermore, if $H=0$ then the projection
operator takes the simple form
\begin{equation}
   P_\pm = \frac{1}{2}\left( 1 \pm \frac{1}{6!\sqrt{g}}\e^{m_1\ldots m_6}
          \pt_{m_1}X^{I_1}\dots\pt_{m_6}X^{I_6}\G_{I_1\ldots I_6} \right)
\end{equation}
where $\s^m$, $m=0,\dots ,5$ label the coordinates on the five-brane
and $g$ is the determinant of the induced metric
\begin{equation}
   g_{mn} = \pt_m X^I \pt_n X^J g_{IJ}\; .
\end{equation}

If the brane configuration is invariant for some combination of
supertranslation $\eta$ and kappa-transformation, then we say it is
supersymmetric. Now $\k$ is a local parameter which can be chosen at
will. Since the projection operators satisfy $P_++P_-=1$, we see that
for a solution of $\d\theta=0$, one is required to set $\kappa=-\eta$,
together with imposing the condition 
\begin{equation}
   P_-\eta = 0
\label{branesusy}
\end{equation} 
For a brane wrapped on a two-cycle in the Calabi--Yau space, spanning
$M_4$ and located at $x^{11}=x$ in the orbifold interval, we
can choose the parameterization 
\begin{equation}
   X^\m = \s^\m \qquad 
   X^A = X^A(\s,\bar{\s}) \qquad
   X^{11} = x
\end{equation}
where $\s=\s^4+i\s^5$. The condition~\eqref{branesusy} then reads 
\begin{equation}
   - \left(i/\sqrt{g}\right) \pt X^A \bar{\pt} X^B \G^{(4)}\G_{AB} \,\eta
        = \eta
\end{equation}
where we have introduced the four-dimensional chirality operator
$\G^{(4)}=\G_0\G_1\G_2\G_3$. Recalling that on the Calabi--Yau three-fold the
Killing spinor satisfies $\G^{\bar{b}}\eta=0$, it is easy to show that
this condition can only be satisfied if the embedding is holomorphic,
that is $X^a=X^a(\s)$, independent of $\bar{\s}$. The condition then
further reduces to 
\begin{equation}
   \G^{(4)}\eta = i \eta
\label{4dchiral}
\end{equation}
which, given that the spinor has definite chirality in eleven dimensions as
well as on the Calabi--Yau space, implies that $\G^{11}\eta=\eta$,
compatible with the global orbifold quotient condition. Thus, finally,
we see that only half of the eight Killing spinors, namely those
satisfying~\eqref{4dchiral}, lead to preserved supersymmetries on the
five-brane. Consequently the low-energy four-dimensional theory
describing the five-brane dynamics will have $\cN=1$ supersymmetry. 

\vspace{0.4cm}

The simplest excitations on the five-brane surviving in the low-energy
four-dimensional effective theory are the moduli describing the
position of the five-brane in eleven dimensions. There is a single
modulus $X^{11}$ giving the position of the brane in the orbifold
interval. In addition, there is the moduli space of holomorphic curves
$\cC_2$ in $X$ describing the position of the brane in the
Calabi--Yau space. This moduli space is generally complicated, and we will
not address its detailed structure here. (As an example, 
the moduli space of genus one curves in K3 is K3
itself~\cite{vb}.) However, we note that these moduli are scalars in
four dimensions, and we expect them to arrange themselves as a set of chiral
multiplets, with a complex structure presumably inherited from that of
the Calabi--Yau manifold. 

Now let us consider the reduction of the self-dual three-form degrees
of freedom. (Here we are essentially repeating a discussion given
in~\cite{wbranes,KLMVW}.) The holomorphic curve is a Riemann surface
and, so, is characterized by its genus $g$. One recalls that the number
of independent harmonic one-forms on a Riemann surface is given by
$2g$. In addition, there is the harmonic volume two-form
$\Omega$. Thus, if we decompose the five-brane world-volume as
$\cC_2\times M_4$, we can expand $H$ in zero modes as 
\begin{equation}
   H=da\wedge\O+F^u\wedge\l_u+h
\end{equation}
where $\l_u$ are a basis $u=1,\dots ,2g$ of harmonic one-forms on
$\cC_2$, while the four-dimensional fields are a scalar $a$, $2g$
$U(1)$ vector fields $F^u=dA^u$ and a three-form field strength
$h=db$. However, not all these fields are independent
because of the self-duality condition $H=*H$. Rather, one easily concludes
that 
\begin{equation}
   h=*da
\end{equation}
and, hence, that the four-dimensional scalar $a$ and two-form
$b$ describe the same degree of freedom. To analyze the vector
fields, we introduce the matrix ${T_u}^v$ defined by 
\begin{equation}
   *\l_u = {T_u}^v\l_v
\end{equation}
If we choose the basis $\l_u$ such that the moduli space metric
$\int_{\cC_2}\l_u\wedge (*\l_v)$ is the unit matrix, $T$ is antisymmetric and,
of course, $T^2=-1$. The self-duality constraint implies for the
vector fields that
\begin{equation}
 F^u={T_v}^u*F^v\; .
\end{equation}
If we choose a basis for $F^u$ such that
\begin{equation}
 T={\rm diag}\left(\left(\begin{array}{cc}0&1\\-1&0\end{array}\right),\dots ,
   \left(\begin{array}{cc}0&1\\-1&0\end{array}\right)\right)
\end{equation}
with $g$ two by two blocks on the diagonal, one easily concludes that only
$g$ of the $2g$ vector fields are independent. In conclusion, for a genus
$g$ curve $\cC_2$, we have found one scalar and $g$ $U(1)$ vector
fields from the two-form on the five-brane worldvolume. The
scalar has to pair with another scalar to form a chiral $\cN =1$
multiplet. The only other universal scalar available is the zero mode
of the transverse coordinate $X^{11}$ in the orbifold direction. 

Thus, in general, the $\cN=1$ low-energy theory of a single five-brane 
wrapped on a genus $g$ holomorphic curve $\cC_2$ has gauge group $U(1)^g$
with $g$
$U(1)$ vector multiplets and a universal chiral multiplet with bosonic
fields $(a,X^{11})$. Furthermore, there is some number of additional chiral multiplets
describing the moduli space of the curve $\cC_2$ in the
Calabi--Yau three-fold. 

\vspace{0.4cm}

It is well known that when two regions of the five-brane world-volume in
M--theory come into close proximity, new massless states
appear~\cite{wfq,strom}. These are associated with membranes
stretching between the two nearly overlapping five-brane surfaces. 
In general, this can lead to
enhancement of the gauge symmetry. Let us now
consider this possibility, heretofore ignored in our discussion. In
general, one can consider two types of brane degeneracy where parts of
the five-brane world-volumes are in close proximity. The first, and
simplest, is to have $N$ distinct but coincident five-branes, all
wrapping the same cycle $\cC_2$ in the Calabi--Yau space and 
all located at the same
point in the orbifold interval. Here, the new massless states come from
membranes stretching between the distinct five-brane world-volumes. The
second, and more complicated, situation is where there is a degeneracy
of the embedding of a single five-brane, such that parts of the curve
$\cC_2$ become close together in the Calabi--Yau space. 
In this case, the new
states come from membranes stretching between different parts of the
same five-brane world-volume. Let us consider these two possibilities
separately. 

The first case of distinct five-branes is analogous to the M--theory
description of $N$ overlapping type IIB D3-branes, which arise as
$N$ coincident five-branes wrapping the same cycle in a flat torus. In
that case, the $U(1)$ gauge theory on each D3-brane is enhanced to a
$U(N)$ theory describing the full collection of branes. Thus, by
analogy, in our case we would expect a similar enhancement of
each of the $g$ $U(1)$ fields on each five-brane. That is, 
when wrapped on a holomorphic curve
of genus $g$, the full gauge group for the low-energy theory
describing $N$ coincident five-branes becomes $U(N)^g$. 

The second case is in closer analogy to a system considered by Witten
in~\cite{wbranes}. There, for example, a system of two type IIA NS
five-branes intersecting with $g+1$ D4-branes in flat space was lifted
to an M--theory description in terms of a single five-brane with world
volume $\Sigma\times M_4$, where $\Sigma$ is a non-compact Riemann
surface of a particular type. This surface could be completed into a
compact surface $\bar\Sigma$ of genus $g$. In general, $\Sigma$ was
such that the type IIA theory was completely ``Higgsed'', so that the
D4-branes were separated and the gauge symmetry was simply
$U(1)^g$. (One might expect $U(1)^{g+1}$ for $g+1$ D4-branes, but
one of the degrees of freedom is ``frozen out''~\cite{wbranes}.)
Degenerations, where parts of the Riemann surface came close together
in the embedding space, corresponded to overlapping D4-branes and so
led to enhanced gauge symmetry. The exact enhancement depended on the
type of the degeneracy; that is, on how many D4-branes were overlapping.
However, the enhanced gauge group is always of the form of a product
of $U(n)$ and $SU(n)$ groups, such that the total rank of the gauge group,
is $g$. The largest allowed group is $SU(g+1)$. The other
allowed groups correspond to various ``Higgsings'' of $SU(g+1)$ by
fields in the adjoint representation.
This situation is very close to our case, except that
instead of a single five-brane wrapping a non-compact Riemann surface
$\Sigma$ embedded in flat space, we have a single five-brane wrapping
a compact Riemann surface $\cC_2$ embedded in a Calabi--Yau space. We
argued above that, generically, we expect a low-energy theory with
$U(1)^g$ symmetry. This corresponds to the ``fully-Higgsed'' case in
Witten's theory. By direct analogy, we would expect similar
enhancement to products of $U(n)$ and $SU(n)$ groups for degeneracies of the 
embedding such that parts of the curve $\cC_2$ come close together in
the Calabi--Yau three-fold. Again, there would be a constraint on the
total rank being equal to $g$. We might expect that the largest possible
gauge group is similarly $SU(g+1)$. However, understanding the details
of this enhancement requires explicit knowledge of the properties of the
moduli space of holomorphic curves. We will return to this subject
elsewhere \cite{lownext}. 

Summarizing the two cases, we see that for $N$ five-branes wrapping the
same curve $\cC_2$ of genus $g$, we expect that the symmetry is enhanced from
$N$ copies of $U(1)^g$ to $U(N)^g$. Alternatively in the second case,
even for a single brane, we can get enhancement if the embedding
degenerates. In general, $U(1)^g$ enhances to a product of unitary
groups such that the total rank is equal to $g$. The maximal
enhancement is presumably to $SU(g+1)$, and the other allowed groups
correspond to different ``Higgsings'' of $SU(g+1)$ by fields in the
adjoint repesentation. For example, if $g=2$, then $SU(3)$ could be
broken to either $SU(2)\times U(1)$ or $U(1)\times U(1)$. In
all cases, the total rank of the symmetry group is conserved. Finally,
we note that in the case where the Calabi--Yau space itself
degenerates to become a singular orbifold, and the five-branes are
wrapped at the singularity, we could expect more exotic enhancement, in
particular, to gauge groups other than unitary groups. In this paper,
however, we will restrict ourselves to the case of smooth Calabi--Yau
spaces.

\subsection{Low energy effective actions}

Next, we would like to discuss the five-dimensional effective actions that
result from the reduction of Ho\v rava--Witten theory on a background
that includes five-branes.
It has already been explained in section 3.2 how the vacua without
five-branes found in this
paper can be used to construct a sensible five-dimensional theory.
Essentially the same arguments apply here. We begin with the
five-dimensional bulk theory. Clearly, the zero-mode content is
unchanged with respect to the case without five-branes. Thus
we have $\cN =1$ supergravity coupled to $h^{1,1}-1$ vector
multiplets and $h^{2,1}+1$ hypermultiplets. What about the gauging of
the hypermultiplet coset space? Inserting the massless
modes~\eqref{massless} into eq.~\eqref{bser} and calculating $G^{(1)}$
via eq.~\eqref{sol} one finds 
\begin{equation}
 G^{(1)} = \frac{1}{2V}(*\o_{\iz})\sum_{m=0}^n\a^{(m),\iz}\e (z)
\end{equation}
in the interval
\begin{equation}
 z_n\leq |z|\leq z_{n+1}
\end{equation}
for fixed $n$, where $n=0,\dots ,N$, and as in eqn.~\eqref{alpha} we
have introduced the parameters
\begin{equation}
   \a^{(m)}_{i_0} = \frac{\sqrt{2}\e_S}{\rho} \b^{(m)}_{i_0}
\end{equation}
to conform with the notation of \cite{losw1,losw2}.
Hence, we still have a non-zero mode that must be taken into account in the
dimensional reduction. Its form, however, depends on the interval one
is considering. Consequently, the five-dimensional action again contains a
term of the form~\eqref{gauging}, but with $\a_{\iz}$ being replaced by
$\sum_{m=0}^n\a_{\iz}^{(m)}$ for the interval $z_n\leq |z|\leq z_{n+1}$.
In other words, we have gauging in the bulk between each two five-branes
but the gauge charge differs from interval to interval. Since the
bulk potential~\eqref{potential} is directly related to the gauging, it is
subject to a similar replacement of charges. In summary, we conclude that
the bulk theory between any pair of neighboring five-branes in the interval
$z_n\leq |z|\leq z_{n+1}$ is as given in ref.~\cite{losw1,losw2}, but with
$\a_{\iz}$ replaced by $\sum_{m=0}^n\a_{\iz}^{(m)}$.

We now turn to the orbifold planes. They are described by four-dimensional
$\cN =1$ theories at $x^{11}=0,\p\r$ coupled to the bulk. 
The zero mode spectrum on these
planes is, of course, unchanged with respect to the situation without
five-branes. It consists of gauge multiplets corresponding
to the unbroken gauge groups $G^{(1)}$ and $G^{(2)}$, as dictated by
the choice of the internal gauge bundle, and corresponding gauge matter
multiplets. The height of the boundary potentials (see 
eqn.~\eqref{YM}) is now set by the charges $\a_{i_0}^{(0)}$ and
$\a_{i_0}^{(N+1)}$ which, due the presence of additional five-brane
charges, are no longer necessarily equal and opposite. 

Finally, we should consider the worldvolume theories of the three-branes
that originate from wrapping the  five-branes around supersymmetric
cycles. Applying the results of the previous subsection to each of the $N$
five-branes, we have $N$ additional four-dimensional $\cN =1$ theories at
$x^{11}=x_1,\dots ,x_N$ which couple to the five-dimensional bulk. The
field content of such a theory at $x^{11}=x_n$ for $n=1,\dots ,N$ is
generically given by $U(1)^{g_n}$ gauge multiplets, where $g_n$ is the genus
of the holomorphic
curve on which the $n$-th five-brane is wrapped, a universal chiral multiplet
and a number of additional chiral multiplets describing the moduli space
of the holomorphic curve within the Calabi--Yau manifold. By the mechanisms
described at the end of the previous subsection, the $U(1)^{g_n}$ gauge
groups can be enhanced to non-Abelian groups. As the simplest example,
two five-branes located at $x^{11}=x_n$ and $x^{11}=x_{n+1}$ could be
wrapped on the
same Calabi--Yau cycle with genus $g_n$. As long as two five-branes are
separated in the orbifold, that is, $x_{n+1}\neq x_n$, we have two gauge groups
$U(1)^{g_n}$, one group on each brane. However, when the two five-branes coincide, that is, 
for $x_{n+1}=x_n$, these groups are enhanced to
$U(2)^{g_n}$. The precise form of the three-brane world-volume theories
should be obtained by a reduction of the five-brane world-volume theory
on the holomorphic curves, in a target space background of the
undeformed Calabi--Yau space together with the non-zero mode for the
four-form field strength. We leave this to a future 
publication~\cite{lownext}, but note here that we expect those three-brane
theories to have a potential depending on the moduli living on the
three-brane and the projection of the bulk moduli to the three-brane
world-volume. This expectation is in analogy with the theories on the orbifold
planes which, as we have seen, possess such a potential. It has been
shown in ref.~\cite{losw1,losw2} that those boundary potentials provide the
source terms for a BPS double-domain wall solution of the five-dimensional
theory in the absence of additional five-branes. This double domain wall
is the appropriate background for a further reduction to four dimensions.
Again, in analogy, we expect the vacuum of the five-dimensional theory
in the presence of five-branes to be a BPS multi-domain wall. More
precisely, for $N$ five-branes, we expect $N+2$ domain walls with
two world-volumes stretching across the orbifold planes and the remaining
$N$ stretching across the three-brane planes. The r\^ole of the potentials on the
three-brane world-volume theories is to provide the $N$ additional
source terms needed to support such a solution.

Let us finally discuss some consequences for the four-dimensional
effective theory. Clearly, there is a sector of the theory which has just
the conventional field content of four-dimensional $\cN =1$ low-energy
supergravities derived from string theory. More precisely, this is
$h^{1,1}+h^{2,1}$ chiral matter multiplets containing the moduli, gauge multiplets
with gauge group $G^{(1)}\times G^{(2)}\subset E_8\times E_8$ and
corresponding gauge matter. In the presence of five-branes, however,
we have additional sectors of the four-dimensional theory leading
to additional chiral multiplets containing the five-brane moduli and,
even more important, to gauge multiplets with generic gauge group
\begin{equation}
 G=\prod_{n=1}^N U(1)^{g_n}\; .
\end{equation}
At specific points in the five-brane moduli space, one expects enhancement
to a non-Abelian group $G=G_1\times\cdots\times G_M$. As explained above,
in typical cases, the factors $G_m$ can be $U(n)$ and $SU(n)$
groups. We expect the enhancement to preserve the rank, that is, we
have 
\begin{equation}
 {\rm rank}(G)=\sum_{n=1}^{N}g_n\; .
\end{equation}
We recall that $g_n$ is the genus of the curve on which the $n$-th five-brane
is wrapped. As it stands, it appears that the rank could be made arbitrarily
large. However, for a given Calabi--Yau space, we expect a constraint on
the rank which originates from positivity constraints in the the
zero-cohomology condition~\eqref{ccon}. This will be further explored
in \cite{lownext}. As is, the five-brane sectors and the conventional
sector of the theory only interact via the bulk supergravity
fields. Therefore, at this point, they are most naturally interpreted
as hidden sectors. 

We should, however, point out that the presence of five-branes provides
considerably more flexibility in the choice of $G^{(1)}\times G^{(2)}$,
the ``conventional'' gauge group that originates from the heterotic
$E_8\times E_8$.
This happens because it is much simpler to satisfy the zero cohomology
condition~\eqref{ccon} in the presence of five-branes. Let us give an
an example which is illuminating, although not necessary physically
relevant. Consider a Calabi--Yau space $X$ with topologically nontrivial
$\tr R\wedge R$. In addition, we set both $E_8$ gauge field backgrounds to zero, which
implies that the unbroken gauge group is simply $E_8\times E_8$.
Without five-branes, such a background is inconsistent since it is in
conflict with the zero-cohomology condition~\eqref{ccon}. However, if
for each independent four-cycle $\cC_{4i_0}$, we can introduce
$N_{i_0}$ five-branes, each having unit intersection number with the
cycle $\cC_{4i_0}$, such that 
\begin{equation}
 N_{i_0}=-\frac{1}{16\p^2}\int_{\cC_{4i_0}}\tr R\wedge R
\end{equation}
then the zero-cohomology condition is satisfied. Of course, the gauge group
will then be enlarged to $E_8\times E_8\times G$ where the gauge group $G$
originates from the five-branes, as discussed above.

What about the form of the four-dimensional effective action? We have
seen that non-standard embedding without five-branes does not change the
form of the effective action with respect to the standard embedding case.
This could be understood as a direct
consequence of the fact that the five-dimensional effective theory
remains unchanged. Above we have seen, however, that the five-dimensional
effective theory does change in the presence of five-branes. In particular,
its vacuum BPS solution is now a multi-domain wall, as opposed to a
double-domain wall in the case without five-branes. Hence, we expect
the four-dimensional theory obtained as a reduction on this multi-domain
wall to change as well. Let us, as an example of this, calculate the
gauge kinetic functions in four dimensions to linear order in $\e_S$.
Here, we will not do this using the five-dimensional theory but,
equivalently, reduce directly from eleven to four dimensions. We define the
modulus $R$ for the orbifold radius by
\begin{equation}
 R=\frac{1}{2V\p\r}\int_{S^1/Z_2\times X}\sqrt{^7g}\; .
\end{equation}
Note that with this definition, $R$ measures the averaged orbifold size
in units of $2\p\r$. Let us also introduce the $(1,1)$ moduli $a^{i_0}$
in the usual way as
\begin{equation}
 \o_{AB}=a^{\iz}\o_{\iz AB}\; .
\end{equation}
Then, the real parts of the low energy fields $S$ and $T^i$ are given by
\begin{equation}
 {\rm Re}(S)=V\; ,\qquad {\rm Re}(T^{\iz})=VR^{-1}a^{\iz}\; .
\end{equation}
We stress that with these definitions, $S$ and $T^{\iz}$ have the standard
K\"ahler potential, that is, the order $\e_S$ corrections to the
K\"ahler potential vanish~\cite{low1}.
The gauge kinetic functions can be directly read off from the 10--dimensional
Yang--Mills actions~\eqref{YM}. Using the metric from eq.~\eqref{sol}
with \eqref{massless}, \eqref{bser} inserted and the above definition of
the moduli, we find
\bea
 f^{(1)} &=& S + \p\e_S T^{\iz}\sum_{n=0}^{N+1}(1-z_n)^2
             \b_{\iz}^{(n)} \\
 f^{(2)} &=& S + \p\e_S T^{\iz}\sum_{n=1}^{N+1}{z_n}^2
             \b_{\iz}^{(n)} \; ,
\eea
where, in addition, we have the cohomology constraint~\eqref{coh1}.
Recall from eq.~\eqref{gkf} that in the case without five-branes, the
threshold correction on the two orbifold planes are identical but opposite
in sign. Note that here the expressions for these two thresholds are,
in fact, different. If, for example, there is only
one five-brane with charges $\b_{\iz}^{(1)}$ at $z=z_1$ on the orbifold, we
have 
\begin{equation}
 f^{(1)}-f^{(2)} = 2\p\e_S T^{\iz}\left[\b_{\iz}^{(0)}+(1-z_1)
                   \b_{\iz}^{(1)}\right]\; .
\end{equation}
We see that the gauge thresholds on the orbifold planes depend on both
the position and the charges of the additional five-branes in the bulk.
This gives considerably more freedom than in the case without five-branes.
In particular, for special choices of the charges and the five-brane
position, the difference of the gauge kinetic functions can be small.
Thus, for instance, the hidden gauge coupling at the physical point
need not be as large as it was in the case without five-branes.

\vspace{0.4cm}

{\bf Note added}
When this manuscript was in preparation we received ref.~\cite{lpt}
which also discusses non-standard embeddings in heterotic M--theory,
however, without considering vacua with five-branes. It also included
an interesting discussion of the appearance of anomalous $U(1)$ gauge
fields. 

\vspace{0.4cm}

{\bf Acknowledgments} 
A.~L.~would like to thank Douglas Smith for helpful discussions and
Graham Ross for hospitality during his stay at Oxford University.
A.~L.~ would also like to thank the Albert--Einstein--Institut for
financial support and Dieter L\"ust for hospitality. We are grateful to
Ron Donagi for helpful email conversations. A.~L.~and B.~A.~O.~are
supported in part by DOE under contract No.~DE-AC02-76-ER-03071. D.~W.~is
supported in part by DOE under contract No. DE-FG02-91ER40671.



\end{document}